# Cyclotron resonance photoconductivity of a two-dimensional electron gas in HgTe quantum wells

Ze-Don Kvon,[a,b*] Sergey N. Danilov,[a] Nikolay N. Mikhailov,[b] Sergey A. Dvoretsky,[b] Wilhelm Prettl,[a] Sergey D. Ganichev[a]

[a]*Terahertz Center, University of Regensburg, Universitätsstraße 31, 93040 Regensburg, Germany*

[b]*Institute of Semiconductor Physics, pr. Lavrentjeva 13, 630090 Novosibirsk, Russia*

**Abstract**

Far-infrared cyclotron resonance photoconductivity (CRP) is investigated in HgTe quantum wells (QWs) of various widths grown on (013) oriented GaAs substrates. It is shown that CRP is caused by the heating of two-dimensional electron gas (2DEG). From the resonance magnetic field strength effective masses and their dependence on the carrier concentration is obtained. We found that the effective mass in each sample slightly increases from the value $(0.0260 \pm 0.0005)m_0$ at $N_s = 2.2 \cdot 10^{11}$ cm$^{-2}$ to $(0.0335 \pm 0.0005)m_0$ at $N_s = 9.6 \cdot 10^{11}$ cm$^{-2}$. Compared to determination of effective masses by the temperature dependence of magnitudes of the Shubnikov-de Haas (SdH) oscillations used so far in this material our measurements demonstrate that the CRP provides a more accurate (about few percents) tool. Combining optical methods with transport measurements we found that the transport time substantially exceeds the cyclotron resonance lifetime as well as the quantum lifetime which is the shortest.

## 1. Introduction

Owing to the advances of molecular beam epitaxy (MBE) technology of narrow gap semiconductors high mobility HgTe quantum wells have recently become available for a wide range of experimental investigations. The two-dimensional electron gas in HgTe is characterized by a highly specific energy spectrum with an inverted band structure, low effective mass $m_n = (0.02 - 0.04)\,m_0$ and large spin splitting at zero magnetic field [1-4]. So far the effective mass in such QW structures has been determined by measurement of SdH oscillations. By this method the effective mass is extracted from the temperature dependence of the SdH amplitudes and is rather indirect. In particularly at low electron densities less than $3 \cdot 10^{11}$ cm$^{-2}$ the uncertainty of SdH oscillations method exceeds 30%-50% [1]. Here we report on the first experimental study of the cyclotron resonance photoconductivity of a 2DEG in HgTe QWs. CRP is the standard method of effective mass determination and gives an accuracy of about few percent.

## 2. Experimental results and discussion

The experiments are carried out on $Cd_{0.7}Hg_{0.3}Te/HgTe/Cd_{0.7}Hg_{0.3}Te$ quantum wells having three different widths: 8 nm, 16 nm and 21 nm. Structures are grown on a GaAs substrate with surface orientation

---

[*] Corresponding author. Institute of Semiconductor Physics, pr. Lavrentjeva 13, 630090 Novosibirsk, Russia Tel.: +7-383-330-6733; fax: +7-383-333-2771; e-mail: kvon@thermo.isp.nsc.ru.

(013) by means of a modified MBE method [5]. Samples with sheet density of electrons $N_s$ from $2 \cdot 10^{11}$ cm$^{-2}$ to $9.6 \cdot 10^{11}$ cm$^{-2}$ and mobility $\mu$ ranging from $10^5$ cm$^2$/Vs to $5 \cdot 10^5$ cm$^2$/Vs have been studied in the temperature range from 2 K to 40 K. The samples are prepared as Hall bars with 50 µm width and 100 µm and 250 µm distances between opposite contacts. For optical excitation we use a molecular terahertz laser with methanol as an active gas optically pumped by a CO$_2$ laser. We use radiation at wavelength 118.8 µm (corresponding to photon energy $\hbar\omega$ = 10.4 meV) with power of about 30 mW. Photoconductivity $\Delta G_{ph}$ has been measured applying standard modulation techniques at constant bias current of a few µA. All samples are also characterized by magnetotransport measurements. As at liquid helium temperature the cyclotron resonance induced photoconductivity is hidden by another photoconductive effect caused by suppression of SdH effect due to electron gas heating, the CRP measurements are carried out at temperature $T > 20$ K where the SdH effect in the photoconductivity is negligible.

The sharp cyclotron resonance peaks in the photoconductivity are detected at the magnetic field $B$ ranging from 2.3 T to 3 T depending on the QW width as well as on the electron density. Figure 1 shows the magnetic field dependence $\Delta G_{ph}(B)$ of the cyclotron resonance induced photoconductivity for all three QW widths.

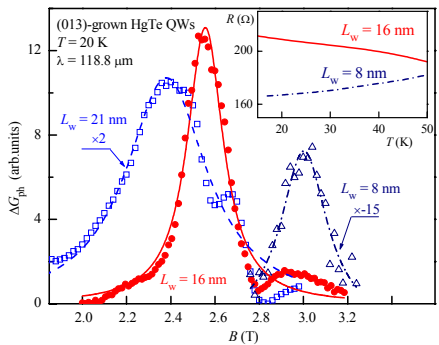

Fig. 1. Photoconductive response to 118.8 µm wavelength of radiation measured in cyclotron resonance geometry as a function of the magnetic field for three HgTe QWs of different thicknesses. Lines are Lorentz-fits with half-width 0.44 T, 0.19 T, and 0.15 T for QWs with 21 nm, 16 nm, and 8 nm width, respectively. The inset shows resistance as a function of temperature.

First we compare the photoconductivity of the sample having 16 nm QW width and electron density $N_s = 3.4 \cdot 10^{11}$ cm$^{-2}$ to the sample of 8 nm width and $N_s = 9.6 \cdot 10^{11}$ cm$^{-2}$. Both resonances can reasonably be approximated by a symmetric Lorentzian with half-width of about 0.15 T ($\approx$ 0.5 meV). However, the sign of the photoconductivity signals is different: while the illumination of the 16 nm QWs results in positive RPC, the RPC of the 8 nm QW is negative. We emphasize that in Fig. 1 the photoconductive signal of the 8 nm QW is inverted. This fact demonstrates that the photoconductivity in our experiments is caused by radiation heating of the electron gas which results in the change of sample resistance. Indeed, as shown in the inset of Fig. 1, these two samples have opposite slopes of the resistance as a function of temperature. The difference in the signal magnitude can be attributed to the difference in slope magnitudes.

Another difference of RPC in the two samples is the position of the resonance peak (see Fig. 1). The half-widths $\Gamma_{CRP}$ of the resonances are almost the same: $\Delta \approx 0.19$ T (0.76 meV) for 16 nm QWs and $\Delta \approx 0.15$ T (0.52 meV) for 8 nm QWs. This is in spite of the fact that the mobility in the 16 nm QWs, being of $4 \cdot 10^5$ cm$^2$/Vs, is four times larger than that of the 8 nm QWs. Comparing $\Gamma_{CRP}$ with the collision broadening $\Gamma_c = \hbar/\tau_{tr}$ obtained from mobility measurements, we find that it is always larger. Indeed for samples with the smaller mobility (8 nm QWs) the $\Gamma_c \approx 0.3$ meV. The experimental observations show that the broadening of the CRP resonance in our samples cannot be due to collision. A possible other broadening mechanism is controlled by the quantum life time $\tau_q$ which determines SdH amplitudes. The quantum lifetime is given by $\tau_q = \hbar / 2\pi k_B T_D$ where $T_D$ is the Dingle temperature which is related to the amplitude $A(B,T)$ of SdH oscillations through

$$A(B,T) = 4\exp\left(-2\pi^2 k_B T_D / \hbar\omega_c\right) \frac{(2\pi^2 k_B T / \hbar\omega_c)}{\sinh(2\pi^2 k_B T / \hbar\omega_c)},$$

where $\omega_c$ is the cyclotron frequency. We obtain from SdH oscillations measured in the same samples the $\Gamma_q \approx (5 - 10)$ meV which is much larger than $\Gamma_{CRP}$ and larger than $\Gamma_c$. In 2DEG heterostructures with dominating small angle scattering $\Gamma_q$ is usually larger than $\Gamma_c$ [6,7], but the obtained here hierarchy of relaxation times as $\tau_{tr} \gg \tau_{CR} \gg \tau_q$ is not typical. We note that most recently the analogous discrepancy between $\tau_{tr}$, $\tau_q$ and $\tau_{CR}$ has been observed in a completely other system: 2DEG in AlGaN/GaN heterostructures [8]. Thus our result shows that this interesting situation can be presented in various 2DEG heterostructures. It is probably due to a nontrivial structure of scattering which is a complicated mixture of long range and short range scattering potentials [9,10].

Now we discuss the CRP results in the third, widest QW of 21 nm width. The main difference in comparison to the previous narrower samples is the presence of shoulders on the high-magnetic field side of the resonance (see Fig. 1). The structure at 2.7 T is displaced by about 0.3 T (1.3 meV) from the main peak. This value is too small for Zeeman spin splitting, which at these fields is about 5 meV. It might be due to zero magnetic field spin splitting caused by inversion asymmetry of the structure [2,4], but to make this conclusion further experiments on structures with different asymmetries are required.

Finally, we obtained cyclotron resonance effective mass as a function of the density of 2D electrons and the width of QWs. In Fig. 2 we show the dependence of the cyclotron effective mass on the electron sheet density in comparison to the SdH effective mass taken from [1].

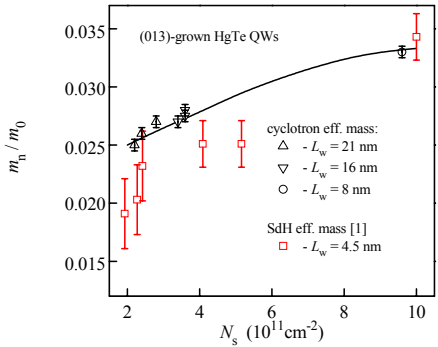

Fig. 2. Effective mass as a function of electron sheet density. Triangles and dot: cyclotron resonance effective mass. Squares: effective masses obtained from SdH measurements taken from [1]. The solid line is a guide for the eye.

It is seen that the effective mass $m_n$ slightly increases from the value $(0.0260 \pm 0.0005)\,m_0$ at $N_s = 2.2\cdot10^{11}$ cm$^{-2}$ to $(0.0335 \pm 0.0005)\,m_0$ at $N_s = 9.6\cdot10^{11}$ cm$^{-2}$. The cyclotron resonance mass, obtained with an accuracy of a few percent, indicates significantly larger values of $m_n$ and much less non-parabolicity than shown by earlier studies and calculations which gave the value of $m_n < 0.02\,m_0$ varying by factor of two in the range of electron densities of Fig. 2.

In summary, cyclotron resonance induced photoconductivity in novel narrow-gap and low-dimensional structures based on HgTe is investigated and cyclotron effective masses are obtained and compared to previous SdH measurements. From cyclotron resonance line widths and magneto-transport measurements relaxation times have been obtained yielding an unusual sequence of times with respect of their magnitude which has only been observed in 2DEG.

**Acknowledgement**

The financial support of the DFG, RFBR (02-05-16591), and programs of RAS "Quantum nanostructures" and "Quantum macrophysics" is gratefully acknowledged.

**References**

[1] A. Pfeuffer-Jeschke et al., Physica B, 256-258 (1998) 486.
[2] X.C. Zhang et al., Phys. Rev. B 63 (2001) 245305.
[3] X.C. Zhang et al., Phys. Rev. B 69 (2004) 115340.
[4] Y.S. Gui et al., Phys. Rev. B 70 (2004) 115328.
[5] V.S. Varavin et al., Proceedings SPIE, 5136 (2003) 381.
[6] M. A. Paalanen, D. C. Tsui, and J. C. Hwang, Phys. Rev. Lett. 51, 2226 (1983).
[7] S. Das Sarma and Frank Stern, Phys. Rev. B 32, 8442 (1985)
[8] S. Syeda et al., Appl. Phys. Lett. 84 (2004) 1507.
[9] E.B. Olshanetsky et al., Phys. Rev. B 68 (2003) 085304.
[10] Hyun-Ick Cho et al., Phys. Rev. B 71 (2005) 245323.